\documentstyle[preprint,version2,aps]{revtex}

\begin{document}
\draft
\begin{title}
\vskip2.5cm
Microscopic Calculations of the Finite-Size Spectrum \\
in the Kondo Problem
\end{title}
\vskip1cm
\author{Satoshi Fujimoto and Norio Kawakami}
\begin{instit}
Yukawa Institute for Theoretical Physics,
Kyoto University, Kyoto 606, Japan
\end{instit}
\author{Sung-Kil Yang}
\begin{instit}
Institute of Physics, University of Tsukuba, Ibaraki 305, Japan
\end{instit}

\begin{abstract}
The finite-size spectrum in the Kondo problem is obtained from
the Bethe-ansatz solution of the exactly solved models.
We investigate the Anderson model,
the highly correlated SU($\nu$) Anderson model and the
{\it s-d} exchange model. For all these models we find that the
spectra exhibit
the properties characteristic of the Fermi liquid fixed point,
and hence our
microscopic calculations are in accordance with the results
obtained by
boundary conformal field theory with current algebra symmetry.
\end{abstract}
\vspace{1 cm}
\narrowtext
\section{Introduction}

The Kondo problem for dilute magnetic impurities in
metals has attracted continuous interest for years.
Many fascinating aspects of the problem have been revealed by
various methods such as the renormalization group
approach \cite{rg}, the local Fermi liquid theory
\cite{fermi,yamada} and the
Bethe-ansatz solution \cite{exact,tw}.
Recently a series of works by Affleck and Ludwig has shed
a new light on
our understanding of the Kondo effect \cite{af1,af2,af3}.
Their new machinery is a technique
of two-dimensional boundary conformal field theories (CFT)
\cite{cardy}.
Various critical properties of the Kondo problem including
the multichannel
overscreened as well as the underscreened case have been
studied
extensively.

The first crucial step made by Affleck in this approach
was to recognize
that for the ordinary Kondo problem the impurity effect
is incorporated as
modifying boundary conditions on the otherwise free
conduction electrons
in the low-energy effective theory \cite{af1}.
{}From the CFT viewpoint,
therefore, the impurity effect should be explicitly
observed in the
finite-size spectrum of the Kondo Hamiltonian.
Assuming that the low-energy
effective Hamiltonian is expressed in the Sugawara form
of the current
algebra Affleck and Ludwig verified the finite-size spectrum
making use of the fusion rule hypothesis \cite{af2}.
On the other hand, the Bethe-ansatz
solutions to various models for the Kondo problem have been
obtained\cite{exact,tw}.
Thus the finite-size
spectrum should be directly obtained starting with the
{\it microscopic
models}. It is then worth
comparing the results with those predicted by boundary CFT.

In this paper we address ourselves to this task and
derive the finite-size spectrum
in the Kondo problem from the Bethe-ansatz solution of
variants of impurity models.
We will show that the essential properties of the
spectrum predicted by boundary CFT indeed agree
with the exact results deduced by the Bethe-ansatz method.
In Sec.II we begin with the Anderson model
which is more tractable than the conventional $s$-$d$ exchange
model (Kondo model) to calculate the finite-size spectrum.
We see that to the first order in $1/L$ with $L$ being the
system size the excitation spectrum coincides with
that of free electrons modified by the phase shift
arising from the impurity scattering.
A similar observation was made earlier for the {\it s-d}
exchange model in \cite{af1}.
Applying the finite-size scaling
we then obtain the canonical (integer) exponents
characteristic of the
local Fermi liquid. We also argue that the anomalous
exponents related to
the $X$-ray absorption singularity are read off from
the finite-size spectrum.
On the other hand, the correlation effects due to the impurity
go into the $1/L^2$ piece in the spectrum. Here
the spin and charge susceptibilities for the impurity
govern the excitation spectrum.
In Sec.III our analysis is extended to the SU($\nu$)
Anderson model (degenerate Anderson model).
We shall point out that the excitation spectrum
is written in terms of the Cartan matrix of the
OSp($\nu,1$) Lie superalgebra, which turns out to be responsible
for the canonical exponents inherent in the
local Fermi liquid. In Sec.IV we turn to
the $s$-$d$ exchange model in which the charge degrees of
freedom of the impurity are completely frozen.
We will see that a careful treatment of the charge
degrees of freedom is necessary in this case. Some technical
details are
summarized in Appendix A.

\section{Anderson model}
\subsection{Bethe-ansatz solution}

In this section our purpose is to calculate the finite-size
spectrum of the
Anderson model. We first recapitulate
the Bethe-ansatz solution of the model while emphasizing
the characteristic
properties common to the Bethe-ansatz solutions to exactly
solved models
for the Kondo problem.
The Anderson model for the magnetic impurity
is defined by the Hamiltonian
\begin{equation}
H=\sum_{k,\sigma}\epsilon_{k}c_{k\sigma}^{\dagger}c_{k\sigma}
+V\sum_{k,\sigma}(c_{k,\sigma}^{\dagger}d_{\sigma}+
d_{\sigma}^{\dagger}
c_{k,\sigma})+\epsilon_{d}
\sum_{\sigma}d^{\dagger}_{\sigma}d_{\sigma}
+Ud^{\dagger}_{\uparrow}d_{\uparrow}
d^{\dagger}_{\downarrow}d_{\downarrow},
\label{eqn:haman}
\end{equation}
with standard notations \cite{tw}. The model describes
free conduction electrons  coupled with
correlated $d$-electrons at the impurity via the resonant
hybridization $V$. After reducing the model to
one dimension  with the use of
partial waves, we  linearize the
spectrum of  conduction electrons as $\epsilon_k =vk$ with
$v=1$ near the Fermi point,
which is equivalent to
the assumption of the  constant density
of states. The hopping operator in the kinetic term is
then replaced
by $-i \partial /\partial x$ in the coordinate representation.
These simplifications enable us to apply  the Bethe-ansatz
method
to diagonalize the Hamiltonian \cite{exact,tw}.
Upon diagonalization we are led to
introduce  two types of rapidity variables, $k_{j}$ and
$\Lambda_{\alpha}$,
for the charge and spin degrees of freedom, respectively.
For the Hamiltonian (\ref{eqn:haman}) on a circle of
circumference $L$
the Bethe-ansatz equations were obtained by Wiegmann
\cite{wiegan}, and by
Kawakami and Okiji \cite{kawa}
\begin{equation}
k_{j}L+\delta(k_{j})=2\pi N_{j}-\sum_{\beta=1}^{M}
\theta_{1}(B(k_{j})-\Lambda_{\beta}),
\label{eqn:bethean1}
\end{equation}
\begin{equation}
2\pi J_{\alpha}+\sum_{\beta=1}^{M}\theta_{2}
(\Lambda_{\alpha}-\Lambda_{\beta})+\sum_{j=1}^{N-2M}
\theta_{1}(\Lambda_{\alpha}-B(k_{j}))
=-2 {\rm Re} \tilde k(\Lambda_{\alpha}) L-2{\rm Re}
\delta(\tilde k(\Lambda_{\alpha})),
\label{eqn:bethean2}
\end{equation}
where $\theta_{n}=2\tan^{-1}(2x/n)$ and
$\delta(k)=-2\tan^{-1}(V^{2}/(k-\epsilon_{d}))$
is the bare phase shift due to the potential scattering
by the impurity at $U=0$, which gives rise to the  resonance
width
$\Gamma=V^2/2$. The total number of (up-spin) electrons is
denoted as
$N~(M)$. Here $\tilde k(\Lambda_\alpha)$
is the complex solution with Im$\tilde k>0$ of the charge
rapidity which
satisfies $iB(\tilde k)-i\Lambda-U \Gamma=0$ where
\begin{equation}
B(k)=k(k-U-2\epsilon_d).
\end{equation}
Comparing Eqs.(\ref{eqn:bethean1}) and (\ref{eqn:bethean2})
with the ordinary Bethe-ansatz
equations we observe
additional phase shift terms $\delta(k)$ proportional
to $1/L$ whose existence is indeed quite relevant to
describe the
physical properties of the impurity part.

\subsection{Ground-state energy}

The exact ground-state energy of the Anderson model is given by
\cite{wiegan,kawa,tw}
\begin{equation}
E_0= \sum_{j=1}^{N-2M} k_j+2\sum_{\alpha=1}^M x(\Lambda_\alpha),
\label{eqn:ground}
\end{equation}
where
\begin{equation}
x(\Lambda)=\epsilon_d+U/2-\Big[ \Lambda+(\epsilon_d+U/2)^2+
\big[ \big( \Lambda+(\epsilon_d+U/2)^2 \big)^2+U^2\Gamma^2/4
\big]^{1/2}
\Big]^{1/2}.
\end{equation}
Note that this expression is divergent because of the constant
density of
states with charge rapidities $k_j$ distributed over
$[-\infty, \beta]$
in the thermodynamic limit.
In order to evaluate the finite-size corrections to the
ground-state energy,
therefore, we have to regularize the sum (\ref{eqn:ground})
upon
taking the large-$L$ limit while keeping the $1/L$
scaling term which is
expected to be universal. Most naively one may cut off
the $k$-integration;
$k \in [-D, \beta]$ with $D$ being the bandwidth. This
prescription,
however, is not suitable for our purpose since introducing
the explicit
momentum cutoff in this way would break conformal invariance.

Instead we are led to introduce a smooth cutoff function
$\varphi (k)$
which is unity near the Fermi point and behaves as
$\vert k\vert^{-(2+\epsilon)}~~
(\epsilon >0)$ when $\vert k\vert \rightarrow \infty$.
Eq.(\ref{eqn:ground}) is
now replaced by
\begin{equation}
E_0= \sum_{j=1}^{N-2M} \varphi (k_j)k_j
+2\sum_{\alpha=1}^M \varphi(x(\Lambda_\alpha)) x(\Lambda_\alpha).
\label{eqn:regulated}
\end{equation}
We apply the Euler-Maclaurin summation formula to
Eqs.(\ref{eqn:bethean1}),
(\ref{eqn:bethean2}) and (\ref{eqn:ground}) to evaluate
the finite-size corrections. The details of
computations are relegated to Appendix A. The final result
is given by
\begin{equation}
E_0=L \varepsilon_0-\frac{\pi v_c}{12L}-\frac{\pi v_s}{12L}~,
\label{eqn:andcasimir}
\end{equation}
where $v_c$ and $v_s$ are the velocities of massless charge
and spin
excitations. It is shown that these velocities take the
same value
$v$ which is equal to unity in our convention (see Appendix A).
To compare the result
(\ref{eqn:andcasimir}) with the finite-size scaling law
in CFT \cite{cardy2} one has to
replace $L$ with $2l$ since $L$ has been defined as the
periodic length of
the system. Then we find the scaling behavior of boundary
CFT \cite{cardy2}
for the
charge and spin sectors, respectively, but with the same
Virasoro central
charge $c=1$.

\subsection{Excitation energy}

Our next task is to calculate the finite-size corrections
to the excitation
energies. For this it is sufficient to take the thermodynamic
limit of
the Bethe-ansatz solution.
Introducing  density functions $\rho(k)$ and $\sigma(\Lambda)$
for the charge and spin rapidities, respectively, we write
the Bethe-ansatz equations in the thermodynamic limit.
Standard calculations yield \cite{wiegan,kawa,tw}
\begin{equation}
\rho (k)={1\over 2\pi}+{1\over L}\Delta (k)+
B^{'}(k)\int^{\infty}_{\alpha}{\rm d}\Lambda a_1(B(k)-\Lambda)
\sigma(\Lambda),
\label{eqn:be1}
\end{equation}
\begin{equation}
\sigma(\Lambda)+\int_{\alpha}^{\infty}{\rm d}\Lambda^{'}
a_2(\Lambda-\Lambda^{'})\sigma
(\Lambda^{'})+\int_{-\infty}^{\beta}{\rm d}k
a_1(\Lambda -B(k))\rho
(k)
=A(\Lambda)+{1\over L}Z(\Lambda),
\label{eqn:be2}
\end{equation}
where
\begin{equation}
a_n (x)=\frac{1}{\pi}\frac{nU\Gamma}{x^2+(nU\Gamma)^2},
\end{equation}
\begin{equation}
A(\Lambda)
=\int^{\infty}_{-\infty} {\rm d}k a_1(B(k)-\Lambda),
\end{equation}
\begin{equation}
Z(\Lambda)
=\int^{\infty}_{-\infty} {\rm d}k a_1(B(k)-\Lambda) \Delta(k),
\end{equation}
and $\Delta (k)=(\Gamma/\pi)[(k-\epsilon_d)^2+\Gamma^2]^{-1}$.
It is clear from Eqs.(\ref{eqn:be1}), (\ref{eqn:be2}) that the
density
functions can be decomposed as
$\sigma=\sigma_h+ (1/L) \sigma_{imp}$
and $\rho=\rho_h+ (1/L) \rho_{imp}$ corresponding to the
host and impurity contributions.
For later convenience let us rewrite Eq.(\ref{eqn:be2})
in such a way that the $k$-integral is taken over the region
$[\beta,\infty]$ instead of $[-\infty, \beta]$.
This can be easily performed by
the Fourier transform. Then Eq.(\ref{eqn:be2}) turns out to be
\begin{equation}
\sigma(\Lambda)+\int_{\alpha}^{\infty}{\rm d}\Lambda^{'}
a_2(\Lambda-\Lambda^{'})\sigma
(\Lambda^{'})=\int^{\infty}_{\beta}{\rm d}k
 a_1(\Lambda -B(k))\rho (k).
\label{eqn:be3}
\end{equation}

The number of conduction electrons $N_h$
is expressed in terms of the density functions
of the host part
\begin{equation}
\frac{N_h}{L}={\cal N}-\biggl(\int^{\infty}_{\beta}{\rm d}k
\rho_h (k)
-2\int^{\infty}_{\alpha}{\rm d}\Lambda \sigma_h (\Lambda)\biggr),
\label{eqn:num1}
\end{equation}
where ${\cal N} =\int_{-\infty}^\infty {\rm d}k \rho_h (k)$ is
an irrelevant
infinite constant.
Similarly for the number of down-spin electrons $M_h$ we get
\begin{equation}
\frac{M_h}{L}=\int^{\infty}_{\alpha}{\rm d}\Lambda
\sigma_h(\Lambda).
\label{eqn:ds1}
\end{equation}
The total energy has
contributions from both of the host and impurity part
\begin{equation}
E= L \int^{\infty}_{\beta}{\rm d}k\biggl[\frac{1}{2\pi}+
\frac{1}{L}\Delta(k)\biggr]\varepsilon_s(k),\label{eqn:energy1}
\end{equation}
where we have introduced the dressed energies
$\varepsilon_s(k)$ and
$\varepsilon_c(\Lambda)$ satisfying \cite{wiegan,kawa,tw}
\begin{equation}
\varepsilon_s(k)=-\varphi(k)k+\int^{\infty}_{\alpha}{\rm d}
\Lambda
a_1(\Lambda-B(k))\varepsilon_c(\Lambda),
\label{eqn:denes}
\end{equation}
\begin{equation}
\varepsilon_c(\Lambda)=2\epsilon_d+U+\int^{\infty}_{\beta}
{\rm d}k
B^{'}(k)a_1(B(k)-\Lambda)\varepsilon_s(k)
-\int^{\infty}_{\alpha}
{\rm d}\Lambda^{'}a_2(\Lambda-\Lambda^{'})
\varepsilon_c(\Lambda^{'}).
\label{eqn:denec}
\end{equation}

Let us now evaluate the finite-size corrections in the
excitation energy spectrum.
We first expand the total energy
(\ref{eqn:energy1}) to the second order
in the variations $\Delta\alpha$ and $\Delta\beta$ of
the integration limits
$\alpha$ and $\beta$.
The first order terms give a chemical
potential which is usually set to zero.
However, since the shift of the chemical potential
due to the impurity is of order $1/L$
we must incorporate it as the finite-size
corrections to the energy spectrum.
In order to estimate this, it is crucial to
notice the relation \cite{wiegan,kawa,tw}
\begin{equation}
\frac{1}{2\pi}(2\epsilon_d+U)= \int^{\alpha}_{-\infty}{\rm
d}\Lambda\sigma_h(\Lambda),
\label{eqn:edal}
\end{equation}
which holds for the host density function. This relation implies
that the impurity level $\epsilon_d$ will change if we vary
the number of conduction electrons by tuning
$\alpha$ since our convention is to measure the impurity
level $\epsilon_d$
from the Fermi level.
Using Eqs.(\ref{eqn:num1}) and (\ref{eqn:edal}) one obtains
the energy shift due to the impurity
\begin{equation}
\Delta E^{(1)}=
\frac{d E}{d \alpha}\Delta \alpha= -\frac{\pi}{L}
\frac{\partial E}{\partial \epsilon_d}\Delta N_{h}
=-\frac{\pi n_d}{L}\Delta N_{h},
\label{eqn:shift}
\end{equation}
where $n_d$ ($=\sum_\sigma \langle d^{\dag}_\sigma
d_\sigma \rangle$)
is the average number of localized electrons given by
\begin{equation}
n_d=1-\int_{-\infty}^\alpha {\rm d}\Lambda
\sigma_{imp}(\Lambda),
\end{equation}
at the impurity site.
Using the Friedel sum rule $n_d=2\delta_F/\pi$ with $\delta_F$
being
the phase shift at the Fermi level we now find
\begin{equation}
\Delta E^{(1)}=-\frac{2\delta_F}{L}\Delta N_{h}.
\label{eqn:chemical}
\end{equation}

The second order terms are obtained from
Eqs.(\ref{eqn:energy1}) $\sim$ (\ref{eqn:denec}).
The result reads
\begin{equation}
\frac{\Delta E^{(2)}}{L}=
-\sigma(\alpha)
\frac{\partial \varepsilon_c(\Lambda)}{\partial
\alpha}\biggr\vert_{\Lambda=\alpha}
\frac{(\Delta\alpha)^2}{2}
-\rho(\beta)
\frac{\partial \varepsilon_s(k)}{\partial
\beta}\biggl\vert_{k=\beta}
\frac{(\Delta\beta)^2}{2}.
\label{eqn:ensp}
\end{equation}
It is  convenient for our purpose to express $\Delta\beta$
and $\Delta\alpha$
in terms of the change of the number of down-spin electrons and
the magnetization $S_h$ ($=N_h/2-M_h$) of  host electrons.
Taking the derivatives
of Eqs.(\ref{eqn:num1}), (\ref{eqn:ds1}) with respect to
$\alpha$ and $\beta$
we have
\begin{equation}
\frac{2}{L}\frac{\partial S_h}{\partial \beta}=
\rho_h(\beta)-\int_{\beta}^{\infty}{\rm d}k
\frac{\partial \rho_h(k)}{\partial \beta}
=\rho_h(\beta)\xi_{ss}(\beta),
\label{eqn:sb}
\end{equation}
\begin{equation}
\frac{2}{L}\frac{\partial S_h}{\partial \alpha}=
-\int_{\beta}^{\infty}{\rm d}k
\frac{\partial \rho_h(k)}{\partial \alpha}
=\sigma_h(\alpha)\xi_{sc}(\alpha),
\label{eqn:sa}
\end{equation}
\begin{equation}
\frac{1}{L}\frac{\partial M_h}{\partial \beta}=
\int_{\alpha}^{\infty}{\rm d}\Lambda
\frac{\partial \sigma_h(\Lambda)}{\partial \beta}
=-\rho_h(\beta)\xi_{cs}(\beta),
\label{eqn:mb}
\end{equation}
\begin{equation}
\frac{1}{L}\frac{\partial M_h}{\partial \alpha}=
-\sigma_h(\alpha)
+\int_{\alpha}^{\infty}{\rm d}\Lambda
\frac{\partial \sigma_h(\Lambda)}{\partial \alpha}
=-\sigma_h(\alpha)\xi_{cc}(\alpha),
\label{eqn:ma}
\end{equation}
where the $2 \times 2$ dressed charge matrix  $(\xi_{ij})$
is given by the
solutions to the integral equations
\begin{equation}
\xi_{ss}(k)=1+\int_{\alpha}^{\infty}{\rm d}\Lambda
a_1(\Lambda -B(k))\xi_{cc}(\Lambda),
\label{eqn:xi1}
\end{equation}
\begin{equation}
\xi_{sc}(\Lambda)=\int_{\beta}^{\infty}{\rm d}k
B^{'}(k)a_1(B(k)-\Lambda)\xi_{ss}(k)
-\int_{\alpha}^{\infty}{\rm d}\Lambda'
a_2(\Lambda -\Lambda^{'})\xi_{sc}(\Lambda^{'}),
\end{equation}
\begin{equation}
\xi_{cs}(k)=\int_{\alpha}^{\infty}{\rm d}\Lambda a_1(B(k)-\Lambda)
\xi_{cc}(\Lambda),
\end{equation}
\begin{equation}
\xi_{cc}(\Lambda)=1+\int_{\beta}^{\infty}{\rm d}k
B^{'}(k)a_1(B(k)-\Lambda)\xi_{cs}(k)
-\int_{\alpha}^{\infty}{\rm d}\Lambda'
a_2(\Lambda -\Lambda^{'})\xi_{cc}(\Lambda^{'}).
\label{eqn:xi4}
\end{equation}
Consequently $\Delta\alpha$ and $\Delta\beta$ are written
in terms of
$\Delta S_h$ and $\Delta M_h$ which represent the deviations
from the
ground-state values.
{}From Eqs.(\ref{eqn:ensp})$\sim$(\ref{eqn:ma})
we thus obtain
\begin{equation}
\Delta E^{(2)}=
\frac{2\pi v_c}{L}\frac{1}{2}
\biggl(\frac{2\xi_{cs}\Delta S_h+\xi_{ss}\Delta M_h}
{{\rm det}\{\xi_{ij}\}}\biggr)^2
+\frac{2\pi v_s}{L}\frac{1}{2}
\biggl(\frac{2\xi_{cc}\Delta S_h+\xi_{sc}\Delta M_h}
{{\rm det}\{\xi_{ij}\}}\biggr)^2,
\label{eqn:ensp2}
\end{equation}
where the "velocities" of spin and charge excitations
are defined as
\begin{equation}
2\pi v_c=-\biggl[\frac{1}{\sigma_h(\alpha)}+ \frac{1}{L}
\frac{\sigma_{imp}(\alpha)}{\sigma_h(\alpha)^2}\biggr]
\, \frac{\partial\varepsilon_c(\Lambda)}{\partial \alpha}
\biggl\vert_{\Lambda=\alpha}\biggr.,
\end{equation}
\begin{equation}
2\pi v_s=-\biggl[\frac{1}{\rho_h(\beta)} + \frac{1}{L}
\frac{\rho_{imp}(\beta)}{\rho_h(\beta)^2}\biggr]
\, \frac{\partial\varepsilon_s(k)}
{\partial \beta}\biggl\vert_{k=\beta}\biggr..
\end{equation}

For vanishing magnetic fields,  we can solve Eqs.(\ref{eqn:xi1})
$\sim$ (\ref{eqn:xi4})  using
the standard Wiener-Hopf method.  Note first
that the dressed charges in these equations become
independent of
$\alpha$ in the limit of zero magnetic field for which
$\beta \rightarrow -\infty$. Hence we can simply
set $\alpha = -\infty$, which corresponds to
the so-called symmetric model with  $2\epsilon_d+U=0$.
The dressed charge matrix is thus cast in
a simple formula
\begin{equation}
\left(
\begin{array}{cc}
 \xi_{ss}& \xi_{cs} \\
 \xi_{sc}& \xi_{cc}
\end{array}\right)=
\left(
\begin{array}{cc}
 \xi(\beta)& \frac{\xi(\beta)}{2} \\
 0 & \frac{1}{\sqrt{2}}
\end{array}\right),
\end{equation}
with $\beta\rightarrow -\infty$.
Here $\xi(\beta)$ is obtained from the solution to the equation
\begin{equation}
\xi(k)=1+\frac{1}{2\pi}\int^{\infty}_{\beta}{\rm d}k
B^{'}(k)K(B(k)-B(k^{'}))\xi(k^{'}),
\end{equation}
with the kernel given by
\begin{equation}
K(x)=\frac{1}{2U\Gamma}\int^{\infty}_{0}{\rm d}p\frac{{\it
e}^{-p/2}}{\cosh{p/2}}
\cos{\frac{px}{2U\Gamma}}.
\end{equation}
One can readily solve this equation by
the Wiener-Hopf technique in the
limit of $\beta \rightarrow -\infty$, obtaining
\begin{equation}
\left(
\begin{array}{cc}
 \xi_{ss}& \xi_{cs} \\
 \xi_{sc}& \xi_{cc}
\end{array}\right)=
\left(
\begin{array}{cc}
 \sqrt{2}& \frac{1}{\sqrt{2}} \\
 0 & \frac{1}{\sqrt{2}}
\end{array}\right).
\end{equation}
It is remarkable that the dressed charges
obtained here are independent of parameters
such as the Coulomb repulsion and the electron filling.
As we shall see
momentarily this fact
ensures the canonical exponents for conduction electrons
inherent in
the (local) Fermi liquid.

The particle-hole type excitations are readily taken into account.
Let us
define
non-negative integers $n^+_c$ and $n^+_s$ to specify their
contributions
in the charge and spin sectors.
Substituting now the explicit values of the dressed charge into
Eq.(\ref{eqn:ensp2})
we write down the finite-size spectrum in the following form
\begin{equation}
\Delta E^{(2)}=\frac{2\pi v_c}{L}\biggl[\frac{(\Delta N_h)^2}{4}
+n^+_c\biggr]
+\frac{2\pi v_s}{L}[(\Delta S_h)^2 + n^+_s].
\label{eqn:ensp3}
\end{equation}
Furthermore we notice that the ``velocities'' $v_c,~v_s$
can be written in a transparent
form with the use of susceptibilities.
Recall that the host and impurity parts of the
charge and spin susceptibilities are given by \cite{wiegan,kawa,tw}
\begin{equation}
\chi_c^h=-\frac{\partial N_h}{\partial \epsilon_d}
=-4\sigma_h(\alpha)
\biggl(\frac{\partial \varepsilon_c(\Lambda)}
{\partial \alpha}\biggl\vert_{\Lambda=\alpha}\biggr.\biggr)^{-1},
\end{equation}
\begin{equation}
\chi_c^{imp}=-\frac{\partial n_d}{\partial \epsilon_d}
=-4\sigma_{imp}(\alpha)
\biggl(\frac{\partial \varepsilon_c(\Lambda)}
{\partial \alpha}\biggl\vert_{\Lambda=\alpha}\biggr.\biggr)^{-1},
\label{eqn:ccimp}
\end{equation}
\begin{equation}
\chi_s^h=\frac{\partial S_h}{\partial H}
=-\rho_h(\beta)
\biggl(\frac{\partial \varepsilon_s(k)}
{\partial \beta}\biggl\vert_{k=\beta}\biggr.\biggr)^{-1},
\end{equation}
\begin{equation}
\chi_s^{imp}=\frac{\partial s_z}{\partial H}
=-\rho_{imp}(\beta)
\biggl(\frac{\partial \varepsilon_s(k)}
{\partial \beta}\biggl\vert_{k=\beta}\biggr.\biggr)^{-1},
\label{eqn:csimp}
\end{equation}
where the magnetization $s_z$ of the impurity part reads
\begin{equation}
s_z=\frac{1}{2}\int_{-\infty}^\beta {\rm d}k \rho_{imp}(k).
\end{equation}
We find
\begin{equation}
 v_c=v\biggl[1 + \frac{1}{L}
\frac{\chi_{c}^{imp}}{\chi_{c}^{h}}\biggr],
\end{equation}
\begin{equation}
 v_s=v\biggl[1 + \frac{1}{L}
\frac{\chi_{s}^{imp}}{\chi_{s}^{h}}\biggr],
\end{equation}
where the velocity of free electrons is
$v=1$ in the present formulation.

Combining Eqs.(\ref{eqn:chemical}) and (\ref{eqn:ensp3})
we finally arrive at
the simple expression for the finite-size spectrum
\begin{equation}
E=E_0+\Delta E^{(1)}+\Delta E^{(2)}
=E_0+\frac{1}{L}E_1+\frac{1}{L^2}E_2,
\label{eqn:scaling}
\end{equation}
where the terms of order $1/L$ and $1/L^2$ are given
respectively by
\begin{equation}
\frac{1}{L}E_1=\frac{2\pi v}{L}
\biggl[\frac{1}{4}\biggl(\Delta N_{h}-2\frac{\delta_F}{\pi}
\biggr)^2-\biggl(\frac{\delta_F}{\pi}\biggr)^2
+n^{+}_c\biggr]
+\frac{2\pi v}{L}[(\Delta S_{h})^2+n^{+}_s],
\label{eqn:finene1}
\end{equation}
and
\begin{equation}
\frac{1}{L^2}E_2=\frac{2\pi v}{L^2}
\frac{\chi_{c}^{imp}}{\chi_{c}^{h}}
 \biggl[\frac{(\Delta N_h)^2}{4}+n^{+}_c\biggr]
+\frac{2\pi v}{L^2} \frac{\chi_{s}^{imp}}{\chi_{s}^{h}}
[(\Delta S_h)^2+n^{+}_s].
\label{eqn:finene2}
\end{equation}
We remark that the bulk electrons do not contribute to
the $1/L^2$ piece because of the linearized dispersion.
Hence the terms of order $1/L$ and order $1/L^2$ give
the finite-size spectrum of bulk electrons and
impurity electrons,
respectively.

\subsection{Critical exponents and universal relations}

We now wish to discuss our results
(\ref{eqn:scaling})$\sim$(\ref{eqn:finene2}) in
view of the finite-size scaling in CFT. It has already
been found that the
low-energy critical behavior in the Kondo effect is
described by boundary
CFT in which we have only left (or right) moving sector of
CFT \cite{af1,af2,af3}. The $1/L$
formula (\ref{eqn:finene1}) in fact exhibits the scaling
behavior predicted by
boundary CFT. Notice again
that when comparing with CFT formula one has to replace
$L$ with $2l$ as pointed out before.
We observe clearly from Eqs.(\ref{eqn:andcasimir})
and (\ref{eqn:finene1}) that
the charge sector is described by $c=1$ CFT with U(1) symmetry
and the spin sector by
$c=1$ SU(2) Kac-Moody CFT at low energies.

In Eq.(\ref{eqn:finene1}) it is seen that the finite-size
spectrum of the
bulk electron part depends on the
non-universal phase shift $\delta_F$. We should remember
that this constant
phase shift $\delta_F$ is equivalent to the chemical
potential due to the
impurity, see Eq.(\ref{eqn:chemical}).
This implies that the effect of the phase
shift amounts to merely imposing twisted boundary conditions
on conduction electrons \cite{af1,af2,af3}.
When reading off from Eq.(\ref{eqn:finene1})
the dimension of the scaling operator associated with
conduction electrons,
we should thus ignore the $\delta_F$ dependence.
Then the scaling dimension $x$ of the conduction electron field
is determined by choosing the quantum numbers
\begin{equation}
\Delta N_h=1, \hskip 5mm \Delta S_h= 1/2,
\label{eqn:quanta}
\end{equation}
which yields
$x=1/2$ upon ignoring $\delta_F$. This is consistent
with the fact that the
system is described by the strong-coupling fixed point of
the local Fermi liquid \cite{fermi,af1}.

According to boundary CFT \cite{cardy}, however, there
exists a one-to-one
correspondence between energy eigenstates and boundary
scaling operators.
Such operators have scaling dimensions explicitly
dependent of the phase
shift. Our intention now is to identify a boundary
operator which is specified
in particular by a ``single-electron'' quantum number
(\ref{eqn:quanta}).
This type of excitation could be relevant to describe the
long-time asymptotic behavior of the electron correlator
in which the hybridization $V$ in Eq.(\ref{eqn:haman}) is
switched on
suddenly at time $t=0$ and $V(t)=V$ for $t>0$.
The correlation exponent will be obtained by setting
$\Delta N_h=1$ and $\Delta S_h=1/2$ in
Eq.(\ref{eqn:finene1}), but
without ignoring $\delta_F$. We get
\begin{equation}
\eta_s=1-2\delta_F/\pi.
\label{eqn:exp}
\end{equation}
Actually such phenomena
have been well known to occur as the
{\it orthogonal catastrophe} which
is induced by the time-dependent local perturbation,
e.g. a power-law singularity in the $X$-ray absorption
profile \cite{and}.
It is quite remarkable that
the critical exponent for the excitonic correlation part
is precisely given by Eq.(\ref{eqn:exp}) \cite{noz}.
Therefore we observe interesting evidence that boundary
CFT will also
play a role to clarify
the critical behavior related to the orthogonal catastrophe.

We point out another important property
characteristic of the local Fermi liquid,
i.e. the universal relations among the static quantities.
To see this we first recall that in Tomonaga-Luttinger
liquids ($c=1$ CFT)
there exist universal relations among the
velocities  and  the susceptibilities such as
\cite{hal,frahmkorepin,shulz,kawa2}
\begin{equation}
v_c\chi_c= \frac{2K_\rho}{\pi},
\hskip 5mm v_s\chi_s=\frac{1}{2\pi},
\label{eqn:lutrel}
\end{equation}
for zero magnetic fields, where $K_\rho$ is a Gaussian
coupling constant
which features the $U(1)$ conformal critical line
in the charge sector.
An important point is that the above relations
hold in the present case not only for the
conduction electron part but also for the impurity part with
the fixed value $K_\rho=1$ independent of the
interaction.
We can see this from the fact
that the term proportional to  $1/L^2$ of the energy spectrum
(\ref{eqn:finene2}) with $n^{+}_c=n^{+}_s=0$ can be cast
into the expression analogous to the $1/L$ term
\begin{equation}
\Delta E^{(2)}_{imp}=
2\pi v_c^{imp}\frac{(\Delta n_d)^2}{4}
+2\pi v_{s}^{imp}(\Delta s_z)^2.
\label{eqn:fscimp}
\end{equation}
Here we have introduced
the velocities of  the impurity part through
\begin{equation}
v_c^{imp}=\frac{\pi }{3\gamma_c^{imp}},
\hskip 5mm
v_s^{imp}=\frac{\pi }{3\gamma_s^{imp}},
\end{equation}
with $\gamma_c^{imp}$, $\gamma_s^{imp}$ being the
specific heat
coefficients of the charge and spin degrees of freedom
of the impurity
electrons.
This expression is essentially the same as in the bulk.
{}From (\ref{eqn:ccimp}), (\ref{eqn:csimp}) and
(\ref{eqn:fscimp})
it is shown that the universal relations for static quantities
(\ref{eqn:lutrel}) with $K_\rho=1$ hold for the impurity part.
This is nothing but the properties of the local Fermi liquid
\cite{fermi}.

\section{SU($\nu$) Anderson model}

An integrable generalization of the infinite-$U$ Anderson model
with  SU($\nu$) spin degrees of freedom
is given by the Hamiltonian \cite{schlott}
\begin{eqnarray}
H&=&\sum_{m=1}^\nu \int{\rm d}x c^{\dagger}_{m}(x)\Big( -i
\frac{\partial}{\partial x} \Big) c_{m}(x)
+\epsilon_{f}\sum_{m=1}^\nu \vert m\rangle
\langle m\vert \nonumber \\
& &+V\sum_{m=1}^\nu \int{\rm d}x\delta (x)
\Big( \vert m\rangle\langle 0\vert
c_{m}(x)+c_{m}^{\dagger}(x)\vert 0\rangle\langle m\vert \Big),
\label{eqn:hamsun}
\end{eqnarray}
where impurity electrons have $\nu$-fold degeneracy.
$\vert 0\rangle$ and $\vert m\rangle~(m=1,2,\cdots ,\nu)$
represent
the unoccupied and single occupied states of the impurity site,
respectively,
and the double occupancy is forbidden
by infinite repulsive Coulomb
interactions among impurity electrons.
The Hamiltonian (\ref{eqn:hamsun}) is presented in the
coordinate space and the spectrum for
conduction electrons has already been linearized.
This model has often been used
to investigate the Kondo effect for rare-earth impurities
in a metal.

The diagonalization of the SU($\nu$) Anderson model
can be performed by introducing $\nu-1$ spin rapidities
$\Lambda^{(l)}_{\alpha}~ (l=0,1,\cdots,\nu-2)$
besides the charge rapidity $\Lambda^{(\nu-1)}_\alpha$.
According to Schlottmann \cite{schlott}
the Bethe-ansatz equations take the form
\begin{eqnarray}
(l+1)\Lambda^{(l)}_{\alpha}L&=&2\pi I^{(l)}_{\alpha}
-[\pi+2{\rm tan^{-1}}((\Lambda^{(l)}_{\alpha}-
\epsilon_f)/(l+1)\Gamma)]
\\ \nonumber
&&+2\sum^{\nu-1}_{q=0}\sum^{p_{lq}}_{p=0}\sum^{K_q}_{\beta=1}
{\rm tan^{-1}}((\Lambda^{(l)}_{\alpha}-\Lambda^{(q)}_{\beta})/
(l+q-2p)\Gamma),  \\ \nonumber
&& \alpha=1,2,\cdots,K_l, \quad K_l=N_l-N_{l+1},
\quad N_{\nu}=0,\quad
l=0,1,\cdots,\nu-1,
\label{eqn:sunbethe}
\end{eqnarray}
where integers $I^{(l)}_{\alpha}$ are quantum numbers
associated with  the
spin and charge degrees of freedom, and
$N_l$ is the number of electrons with each spin.

We henceforth concentrate on the excitation spectrum.
As mentioned in the previous SU(2) case, we can work with
the  rapidity
density functions $\sigma^{(l)}(\Lambda)$ defined
in the thermodynamic limit when   calculating
the finite-size corrections to excitation energies.
These density functions satisfy
$\nu$-coupled linear equations \cite{schlott}
\begin{eqnarray}
\sigma^{(l)}(\Lambda)=\frac{l+1}{2\pi}
&+&\frac{1}{\pi
L}\frac{(l+1)\Gamma}{(\Lambda-\epsilon_{f})^2
+[(l+1)\Gamma]^2} \\ \nonumber
&-&\sum_{q=0}^{\nu-1}\sum_{p=0}^{p_{lq}}
\int^{B_{q}}_{-\infty}{\rm d}\Lambda^{'}
\sigma^{(q)}(\Lambda^{'})
K_{pq}^{(l)}(\Lambda-\Lambda^{'}), \\ \nonumber
&& \qquad\qquad\qquad\quad l=0,1,\cdots,\nu-1,
\label{eqn:sig1}
\end{eqnarray}
where $p_{lq}={\rm min}(l,q)-\delta_{lq}$ and
$\Gamma=V^2/2 $.
The integral kernel is defined by
\begin{equation}
K_{pq}^{(l)}(x)=\frac{1}{\pi}
\frac{(l+q-2p)\Gamma}{x^2+[(l+q-2p)\Gamma]^2}.
\end{equation}
We note here again that the density functions
are decomposed into the sum of the host part
and the impurity part:
$\sigma^{(l)}=\sigma^{(l)}_h+ (1/L)\sigma^{(l)}_{imp}$.
The total number of
host electrons is then given by
\begin{equation}
\frac{N_h}{L}=\sum^{\nu-1}_{l=0}(l+1)
\int^{B_{l}}_{-\infty}{\rm d}\Lambda
\sigma^{(l)}_h(\Lambda).
\label{eqn:tnum}
\end{equation}
On the basis of these equations we
compute the finite-size spectrum  following the
method explained in the preceding sections.
To proceed with clarity we
will treat the charge and spin sectors
separately.

\subsection{Charge excitation}

Let us first evaluate the finite-size corrections in
the charge sector. For this purpose
we can put $B_l\rightarrow -\infty$ for $l=0,1,\cdots, \nu-2$
while $B_{\nu-1}=Q$, which
corresponds to the case of zero magnetic fields.
The value of $Q$ is determined by the number of
host electrons through
Eq.(\ref{eqn:tnum}).
The total energy including the impurity contribution
is simply expressed as \cite{schlott}
\begin{equation}
\frac{E}{L}=\int^{Q}_{-\infty}{\rm d}\Lambda
\biggl[\frac{\nu}{2\pi} +\frac{1}{\pi L}
\frac{\nu\Gamma}{(\Lambda-\epsilon_{f})^2+
(\nu\Gamma)^2}\biggr]\varepsilon^{(\nu-1)}(\Lambda),
\label{eqn:ene}
\end{equation}
in terms of the  dressed energies
$\varepsilon^{(l)}(\Lambda)~(l=0,1,\cdots ,\nu-1)$
which are defined by the
following equations
\begin{equation}
\varepsilon^{(l)} (\Lambda)=\nu\Lambda\delta_{l, \nu-1}-
\sum_{m=0}^{l}\int_{-\infty}^{Q}{\rm d}
\Lambda^{'}K_{m l}^{(\nu-1)}
(\Lambda-\Lambda^{'})\varepsilon^{(\nu-1)}(\Lambda^{'}).
\end{equation}
The finite-size corrections are computed by
expanding Eq.(\ref{eqn:ene}) in $\Delta Q$ to the
second order.  Similarly to the SU(2) model, the
first order term in $\Delta Q$ arises from the impurity
part.  Expanding the energy to the linear term in $\Delta Q$
we obtain
\begin{equation}
\Delta E^{(1)}=-\frac{\partial E}{\partial Q}\Delta Q
=-\nu\frac{\delta_F}{\pi}\Delta Q,
\label{eqn:dele}
\end{equation}
where we have used the Friedel sum rule \cite{schlott}
\begin{equation}
2\nu\delta_F =2\pi\frac{\partial E}{\partial Q}
=2\pi n_{f},
\end{equation}
with  the average number $n_{f}$ of localized
electrons given by
\begin{equation}
n_f=\nu \int_{-\infty}^Q {\rm \Lambda}
\sigma_{imp}^{(\nu-1)}(\Lambda).
\end{equation}

We next rewrite $\Delta Q$ in terms of
the change of the host electron number $\Delta N_h$.
{}From Eq.(\ref{eqn:tnum}) one finds
\begin{equation}
\frac{1}{L}\frac{\partial N_h}{\partial Q}
=\nu\sigma_h^{(\nu-1)}(Q)
+\nu\int^{Q}_{-\infty}{\rm d}\Lambda
\frac{\partial \sigma_h^{(\nu-1)}(\Lambda)}{\partial Q}=
\sigma_h^{(\nu-1)}(Q)\xi_{\nu-1,\nu-1}(Q),
\label{eqn:devnum}
\end{equation}
where the dressed charge $\xi_{\nu-1,\nu-1}$ is determined by
\begin{equation}
\xi_{\nu-1,\nu-1}(\Lambda)
=\nu-\int^{Q}_{-\infty}{\rm d}\Lambda^{'}
\sum^{\nu-2}_{l=0}K_{l,\nu-1}^{(\nu-1)}(\Lambda-\Lambda^{'})
\xi_{\nu-1,\nu-1}(\Lambda^{'}).
\end{equation}
The Wiener-Hopf solution to this
equation yields $\xi_{\nu-1,\nu-1}(Q)=\sqrt{\nu}$ \cite{schlott}.
Hence we get
\begin{equation}
\Delta Q=\frac{\Delta N_{h}}{\sqrt{\nu}\sigma^{(\nu-1)}_{h}(Q)L}.
\label{eqn:delq}
\end{equation}
Furthermore the host electron density function
$\sigma^{(\nu-1)}_{h}$ at $\Lambda =Q$
is similarly evaluated as
$\sigma^{(\nu-1)}_{h}(Q)=\sqrt{\nu}/2\pi$.
We now obtain the corrections linear in $1/L$
\begin{equation}
\Delta E^{(1)}=-\frac{2\delta_F}{L}\Delta N_{h}.
\label{eqn:dele2}
\end{equation}

Turning to the second order terms we expand
the energy as
\begin{equation}
\frac{\Delta E^{(2)}}{L}=-\sigma^{(\nu-1)}(Q)\frac{\partial
\varepsilon^{(\nu-1)}(\Lambda)}{\partial Q}
\biggl\vert_{\Lambda =Q}
\biggr.\frac{(\Delta Q)^2}{2}.
\label{eqn:enex}
\end{equation}
Using (\ref{eqn:delq}) we  obtain
\begin{equation}
\Delta E^{(2)}=\frac{2\pi v_c}{L}\frac{(\Delta N_h)^2}{2\nu},
\label{eqn:fses}
\end{equation}
where the velocity is defined by
\begin{equation}
2\pi v_c=-\biggl[\frac{1}{\sigma^{(\nu-1)}_h(Q)}
+\frac{1}{L}\frac{\sigma^{(\nu-1)}_{imp}}
{\sigma^{(\nu-1)}_h(Q)^2}\biggr]
\, \frac{\partial \varepsilon^{(\nu-1)}(\Lambda)}{\partial Q}
\biggl\vert_{\Lambda=Q}\biggr. .
\end{equation}
Combining Eqs.(\ref{eqn:fses}) and (\ref{eqn:dele2}),
and repeating similar
manipulations as in the SU(2) case
we obtain  the finite-size spectrum for the charge sector
\begin{equation}
E=E_0+\frac{1}{L}E_1^c+\frac{1}{L^2}E_2^c ~,
\end{equation}
where the terms of order $1/L$ and $1/L^2$ are
given respectively by
\begin{equation}
\frac{1}{L}E_1^c=\frac{2\pi v}{L}
\biggl[\frac{(\Delta N_{h}-\nu\delta_F/\pi)^2}
{2\nu}+n^{+}_c\biggr]
-\frac{\pi v}{L}\nu
\biggl(\frac{\delta_F}{\pi}\biggr)^2,
\label{eqn:fses2}
\end{equation}
and
\begin{equation}
\frac{1}{L^2}E_2^c=\frac{2\pi v}{L^2}
\frac{\chi_{c}^{imp}}{\chi_{c}^h}
\biggl[\frac{(\Delta N_{h})^2}{2\nu}+n^{+}_c\biggr].
\end{equation}
Here the velocity of free electrons is $v=1$ and a
non-negative integer
$n_c^+$ represents the particle-hole excitation in
the charge sector.
Putting $n^{+}_c=0$  we can express the $1/L^2$-order term
in an analogous fashion to the $1/L$-order term
\begin{equation}
\frac{1}{L^2}E_2^c=2\pi v^{imp}_c
\frac{(\Delta n_{f})^2}{2\nu},
\end{equation}
where the velocity of the impurity part is
$v_{c}^{imp}=v\chi_{c}^{h}/\chi_{c}^{imp}$.

\subsection{Spin excitation}

In order to deal with the spin sector we take
the integer-valence limit, $n_{f}\rightarrow 1$
($Q\rightarrow\infty$), to facilitate our calculations.
In this limit
the Bethe-ansatz equations for the spin part reduce to
\cite{schlott}
\begin{eqnarray}
\sigma^{(l)}(\Lambda)=\frac{1}{2\pi}
\frac{\delta_{l,0}}{(\Lambda+1/J)^2+1/4}
-\sum_{q=0}^{\nu-2}\int^{\infty}_{B_{q}}{\rm d}\Lambda
&&A_{lq}(\Lambda -\Lambda^{'})\sigma^{(q)}
(\Lambda^{'}),  \nonumber \\
&&\qquad l=0,1,\cdots,\nu-2,
\label{eqn:spinsig}
\end{eqnarray}
where $J=-\Gamma/\epsilon_{f}$ and
$A_{lq}(\Lambda)$ is the Fourier transform of
\begin{equation}
\tilde{A}_{lq}(x)=\delta_{l,q}(1+e^{-|x|})
-(\delta_{l+1,q}+\delta_{l-1,q})e^{-|x|/2}.
\end{equation}
The values of $B_{l}$'s are determined by
the total number of electrons $N_{l}$ \cite{schlott}.
We have
\begin{equation}
\frac{N_{l}}{L}-\frac{N_{l+1}}{L}
=\int^{\infty}_{B_{l-1}}{\rm d}\Lambda
\sigma^{(l-1)}(\Lambda)-2\int_{B_{l}}^{\infty}{\rm d}\Lambda
\sigma^{(l)}(\Lambda)+\int_{B_{l+1}}^{\infty}{\rm d}\Lambda
\sigma^{(l+1)}(\Lambda),
\end{equation}
for $l=1,2,\cdots, \nu-3$ and
\begin{equation}
\frac{N_0}{L}-\frac{N_{1}}{L}
=1-2\int^{\infty}_{B_0}{\rm d}\Lambda\sigma^{(0)}
(\Lambda)+\int^{\infty}_{B_1}{\rm d}\Lambda\sigma^{(1)}
(\Lambda),
\end{equation}
\begin{equation}
\frac{N_{\nu-2}}{L}-\frac{N_{\nu-1}}{L}
=-2\int^{\infty}_{B_{\nu-2}}{\rm d}
\Lambda\sigma^{(\nu-2)}(\Lambda)
+\int^{\infty}_{B_{\nu-3}}{\rm d}
\Lambda\sigma^{(\nu-3)}(\Lambda).
\end{equation}
We define the dressed energy function
\begin{equation}
\varepsilon^{(l)}=-\nu\Lambda\delta_{l,\nu-2}
-\sum_{q=0}^{\nu-2}\int^{\infty}_{B_{q}}{\rm d}\Lambda^{'}
A_{lq}(\Lambda-\Lambda^{'})\varepsilon^{(q)}(\Lambda^{'}),
\end{equation}
so as to express the total energy in the form
\begin{equation}
\frac{E}{L}=\int^{\infty}_{B_{0}}{\rm d}\Lambda\frac{1}{2\pi}
\frac{1}{(\Lambda +1/J)^2+1/4}\varepsilon^{(0)}(\Lambda).
\label{eqn:totenergy}
\end{equation}
We observe here that Eqs.(\ref{eqn:spinsig})
$\sim$(\ref{eqn:totenergy})
take the form common to integrable  SU($\nu$)
symmetric models
such as the  SU($\nu$) Heisenberg model \cite{izer,devega}
and the
degenerate Hubbard model \cite{frahm}.
Based on the analysis of these
SU($\nu$) models let us
introduce the following quantities
\begin{equation}
\frac{M^{(l)}_h}{L}=\int^{\infty}_{B_{l}}{\rm d}\Lambda
\sigma_h^{(l)}(\Lambda), ~~  \hskip 5mm  l=0,1,\cdots,\nu-2,
\label{eqn:ml}
\end{equation}
as  quantum numbers
characterizing the spin degrees of freedom.
Making use of the SU($\nu$)
results \cite{devega,frahm}
we find the finite-size spectrum of the spin sector
\begin{equation}
E-E_0=\frac{2\pi}{L}\biggl[\frac{1}{2}\Delta {\bf M}^{T}_h
(Z_{\nu}^{-1})^{T}VZ_{\nu}^{-1}\Delta {\bf M}_h
+\sum_{l=0}^{\nu-2}v_ln^{+}_l\biggr]
,\label{eqn:fsess}
\end{equation}
where $\Delta {\bf M}^{T}_h=(\Delta M^{(0)}_h,\cdots,
\Delta M_h^{(\nu-2)})$, $V={\rm diag}(v_{0}, v_{1}, \cdots,
 v_{\nu-2})$ and
$n^{+}_l$ are non-negative integers representing
particle-hole type
excitations in the spin sector.
The dressed charge matrix $Z_\nu$
is defined as $(Z_\nu)_{lm}=\xi_{lm}(B_m)$
where
\begin{equation}
\xi_{lm}(\Lambda)= \delta_{lm}
-\sum_{q=0}^{\nu-2}\int^{\infty}_{B_{q}}{\rm d}\Lambda^{'}
A_{qm}(\Lambda-\Lambda^{'})\xi_{lq}(\Lambda^{'}).
\end{equation}
We note that
the matrix $Z_{\nu}$ enjoys a nice property of
$(Z_{\nu}^{-1})^{T}Z_{\nu}^{-1}$ being
equal to the Cartan matrix for the SU($\nu$) Lie algebra
\cite{devega}.
In Eq.(\ref{eqn:fsess})
the velocities of spin excitations have been defined by
\begin{equation}
v_{l}=-\biggl[\frac{1}{\sigma^{(l)}_h(B_{l})}
+ \frac{1}{L}
\frac{\sigma^{(l)}_{imp}(B_l)}{\sigma^{(l)}_h(B_{l})^2}
\biggr]
\, \frac{\partial\varepsilon^{(l)}(\Lambda)}{\partial B_{l}}
\biggl\vert_{\Lambda=B_{l}},\quad l=0,1,\cdots,\nu-2,
\end{equation}
which are independent of the spin index $l$
for zero magnetic fields.
Consequently the finite-size spectrum for
the spin sector reads
\begin{equation}
E-E_{0}=\frac{2\pi v}{L}\frac{1}{2}\Delta {\bf M}_{h}^{T}
{\cal C_\nu}\Delta {\bf M}_{h}
+\frac{2\pi v^{imp}_{s}}{L^2}\frac{1}{2}(L\Delta {\bf m}^T)
{\cal C_\nu}(L\Delta {\bf m}),
\label{eqn:sunens}
\end{equation}
where we have set $n^{+}_l=0$ for simplicity, ${\cal C_{\nu}}$ is
the $(\nu-1) \times (\nu-1)$ SU($\nu$) Cartan matrix
\begin{equation}
{\cal C_{\nu}}=
\left(
\matrix {2      &   -1    & \null   &
             \smash{\lower1.7ex\hbox{\LARGE 0}}  \cr
         -1     & \ddots  &  \ddots &   \null \cr
        \null   & \ddots  &  \null  &   -1    \cr
        \smash{\hbox{\LARGE 0}}  & \null   &  -1   &   2 \cr}
\right),
\end{equation}
and $\Delta {\bf m}$ denote the variations of
the $(\nu-1)$-component vector ${\bf m}$
defined by
\begin{equation}
m^{(l)}=\int^{\infty}_{B_{l}}{\rm d}\Lambda
\sigma_{imp}^{(l)}(\Lambda),
\end{equation}
for the impurity part.
This completes the calculation of the spin
excitation spectrum.

\subsection{Finite-size spectrum}

We are now ready to present the final result for the
finite-size spectrum.
The $1/L$ piece in the spectrum is obtained by
combining the charge and spin
excitations given by
Eqs.(\ref{eqn:fses2}) and (\ref{eqn:sunens}).
Setting $n_c^+=n^{+}_l=0$ for
simplicity we write the result in the
$\nu \times \nu$ matrix form
\begin{equation}
\frac{1}{L}E_1=
\frac{2\pi v}{L}\frac{1}{2}\Delta {\bf M}^{T}
{\cal C}_{f}\Delta {\bf M}
-\frac{\pi v}{L}\nu
\biggl(\frac{\delta_F}{\pi}\biggr)^2,
\end{equation}
where  $\Delta {\bf M}^T \equiv
(\Delta {\bf M}_h^T, ~\Delta M_h^{(\nu-1)})$
with $\Delta  M_h^{(\nu-1)}=\Delta N_h-\nu\delta_F/\pi$.
Here the $\nu \times \nu$ matrix
${\cal C}_{f}$ is given by
\begin{equation}
{\cal C}_{f}=
\left(
\matrix {2     & -1      & \null  &
                 \smash{\lower1.7ex\hbox{\LARGE 0}} \cr
        -1     & \ddots  & \ddots & \null   \cr
        \null  & \ddots  &  2     & -1      \cr
        \smash{\hbox{\LARGE 0}}   & \null   &  -1    & 1  \cr}
\right) .
\end{equation}
It is amusing that this matrix coincides with
the Cartan matrix for the OSp($\nu,1$) Lie superalgebra.
This matrix
characterizes the energy spectrum of
{\it free electron} systems
as well as the  electron models based on the
OSp($\nu,1$) superalgebra.
That is, the OSp($\nu,1$) Cartan matrix
in the finite-size spectrum gives rise to
{\it canonical} exponents for the correlation functions.
In fact  the critical exponents for
electron correlators
$\langle c_j^{\dag}(t) c_j(0)\rangle \simeq t^{-\eta_n}$
are obtained by specifying the quantum numbers
$\Delta M_h^{(l)}=1$ (or 0)
for $l \geq n$  (or $l < n)$.  We then find  the
critical exponents $\eta_n=1$ irrespective of
$n~(=0,1,\cdots, \nu-1)$.
We see that the finite-size spectrum for the
host part ($1/L$ order) has the form in accordance with
boundary
CFT with U(1) symmetry in the charge sector
and level-1 SU($\nu$)
Kac-Moody symmetry in the spin sector.

On the other hand, the $1/L^2$ part does not
take such a simple form
since the spin and
charge velocities are different from each other
due to the electron correlation effects.
It is worth, however,
considering the hypothetical situation where
these two velocities are equal, say
$v_c^{imp}=v_s^{imp}=1$. Then
the excitation energy is again cast into the form of
the OSp($\nu,1$) Cartan matrix
\begin{equation}
\Delta E^{(2)}=
\frac{2\pi}{L^2} \frac{1}{2} (L\Delta {\bf m}^T)
{\cal C}_{f}(L\Delta {\bf m}).
\label{eqn:sunens2}
\end{equation}
Since the
critical exponents of correlation functions do not
depend on the velocities this result indicates that
the spectrum of
scaling dimensions are solely characterized
by the Cartan matrix.
Furthermore note that the spectrum for the impurity
part ($1/L^2$ order) is essentially given
by the same formula
as that for free electrons {\it except for}
the modification
of velocities.  This result  nicely fits
with the local Fermi-liquid picture of
Nozi\`eres \cite{fermi}.

\section{{\it s-d} exchange model}

This section is devoted to the analysis of the
finite-size spectrum of the
traditional {\it s-d} exchange model. The reader
may be somewhat curious
about why the analysis of the simplest model comes at the end.
As we shall
proceed it is seen that the simplest nature of
the model gives rise
to peculiarity when trying to extract the finite-size
spectrum based on the
Bethe-ansatz solution.

The Hamiltonian of the  {\it s-d} exchange model is given by
\begin{equation}
H=\sum_{\sigma}\int{\rm d}x
c_{\sigma}^{\dagger}(x)\Big( -i\frac{\partial}{\partial x}\Big)
c_{\sigma}(x)
+J\sum_{\sigma\sigma^{'}}{\bf S}\cdot c^{\dagger}_{\sigma}(0)
\bbox{\sigma}_{\sigma\sigma^{'}}c_{\sigma^{'}}(0),
\end{equation}
where conduction electrons are scattered by the
impurity spin located at the origin and
the exchange coupling is assumed to be antiferromagnetic.
Applying the Bethe-ansatz technique Andrei \cite{andrei} and
Wiegmann \cite{wiegsd}
obtained the Bethe-ansatz
equations
\begin{equation}
k_jL=2\pi I_j+\sum^{M}_{\beta=1}\theta(2\Lambda_{\beta}-2),
\quad
j=1,2,\cdots,N_h,
\end{equation}
\begin{equation}
N_h\theta(2\Lambda_{\alpha}-2)+\theta(2\Lambda_{\alpha})
=-2\pi J_{\alpha}+\sum_{\beta=1}^M
\theta(\Lambda_{\alpha}-\Lambda_{\beta}),
\quad \alpha=1,2,\cdots,M,
\end{equation}
where $\theta (x)=-2\tan^{-1}(x/c)$ with
$c=2J/(1-J^2)$, and $N_h$, $M$ are the number of host electrons
and the number of down spins, respectively.
In contrast to the Anderson model
the charge degrees of freedom for the impurity
are completely frozen, so we need a careful
treatment of the charge excitation
in the Bethe-ansatz equations.  A peculiar
aspect in the Bethe equations
is that the charge and spin degrees of freedom are completely
decoupled. Hence a standard technique of the dressed charge
cannot be applied in a straightforward manner
to the charge sector
of the model. For example, a naive calculation of the dressed
charge yields $\xi=1$ for the charge sector,
which is the value for spinless fermions. This seems
apparently  in contradiction to
$\xi=\sqrt{2}$ for free electrons.

Thus we deal with the charge degrees of freedom
directly without relying on the dressed-charge approach.
The total energy $E$ is derived from
the Bethe-ansatz solution \cite{andrei,wiegsd}.
When there are no
magnetic fields we have
\begin{equation}
E=\frac{2\pi}{L} \sum_{j=1}^{N_{h}}n_{j}
+N_{h}\int^{\infty}_{-\infty}{\rm d}
\Lambda [\theta (2\Lambda-2)-\pi]\sigma (\Lambda),
\label{eqn:sdtotalenergy}
\end{equation}
where the density function $\sigma(\Lambda)$ obeys
\begin{equation}
\sigma (\Lambda)=\frac{1}{2\pi}\biggl\{ \frac{N_{h}}{L}
\frac{4c}{(2\Lambda-2)^2+c^2}
+\frac{1}{L}\frac{4c}{(2\Lambda)^2+c^2}
-\int^{\infty}_{B}{\rm d}\Lambda^{'}
\frac{2c}{(\Lambda-\Lambda^{'})^2+c^2}\sigma (\Lambda^{'})
\biggr\},
\label{eqn:sdbethe}
\end{equation}
with $B=-\infty$ for zero magnetic fields.

Let us first examine the charge excitation.
In the limit $B=-\infty$ one
can solve Eq.(\ref{eqn:sdbethe}) using the Fourier transform.
Having obtained
the explicit form of $\sigma (\Lambda)$
it is not difficult to evaluate
the $1/L$ contribution in the charge sector. One finds
\begin{equation}
\frac{1}{L}\Delta E_1^c
= \frac{2\pi v}{L}\frac{1}{4}(\Delta N_{h}-1)^2
-\frac{\pi v}{2L}.
\label{eqn:ench}
\end{equation}
Notice that the shift $\delta N_h \rightarrow \delta N_h -1$
originates from the $\pi/2$ phase shift.

Consider next the spin excitation spectrum with
$N_{h}$ being kept fixed.
The spin part of the total energy in the presence of
a magnetic field is \cite{andrei,wiegsd}
\begin{equation}
E^{s}=N_{h}\int^{\infty}_{B}[\theta (2\Lambda-2)-\pi]
\sigma (\Lambda),
\end{equation}
where $B$ is determined through the
number of down spins $M_{h}$ of host electrons which is
given by
\begin{equation}
{M_{h} \over L}=\int^{\infty}_{B}{\rm d}\Lambda \sigma_h (\Lambda).
\end{equation}
Here the density function is separated
into a host and an impurity part:
$\sigma(\Lambda)=\sigma_h(\Lambda)
+(1/L)\sigma_{imp}(\Lambda)$.
Using the dressed energy defined by
\begin{equation}
\varepsilon (\Lambda)={N_{h} \over L}[\theta (2\Lambda-2)-\pi]
-\frac{1}{2\pi}\int^{\infty}_{B}{\rm d}\Lambda^{'}
\frac{2c}{(\Lambda-\Lambda^{'})^2+c^2}\varepsilon (\Lambda^{'}),
\end{equation}
we can express the total energy as \cite{andrei,wiegsd}
\begin{equation}
E^s = L \int^{\infty}_{B}{\rm d}\Lambda
\biggl\{\frac{N_h}{\pi L}\frac{2c}{(2\Lambda -2)^2+c^2}
+\frac{1}{\pi L}\frac{2c}{(2\Lambda)^2+c^2}\biggr\}
\varepsilon (\Lambda). \label{eqn:sdes}
\end{equation}
The dressed charge is determined by the following equation
\begin{equation}
\xi_{s}(\Lambda)=1-\frac{1}{2\pi}\int^{\infty}_{B}
{\rm d}\Lambda^{'}\frac{2c}{(\Lambda-\Lambda^{'})^2+c^2}
\xi_{s}(\Lambda^{'}),
\end{equation}
which gives $\xi_{s}(B)=1/\sqrt{2}$ reflecting SU(2) symmetry.
The variation $\Delta B$ is expressed
in terms of $\Delta S_h$, i.e.
$\Delta S_h=-\sigma_h (B)\xi_s (B)\Delta B$.
As in the preceding sections,
we then obtain the finite-size spectrum of the spin sector
\begin{equation}
\frac{1}{L}E_1^s+\frac{1}{L^2}E_2^s
=\frac{2\pi v}{L}(\Delta S_{h})^2
+\frac{2\pi v_{imp}}{L^2}(L\Delta s_{z})^2,\label{eqn:sdensp}
\end{equation}
where we have suppressed the particle-hole excitation just for
convenience and the velocity of free fermion $v=1$.
The velocity $v_{imp}$ of spin excitation for
the impurity part
is written in terms of the dressed energy
\begin{equation}
2\pi v_{imp}=\frac{-1}{\sigma_{imp}(B)}
\frac{\partial \varepsilon (\Lambda)}{\partial B}
\biggl\vert_{\Lambda =B}\biggr..
\end{equation}

One can see from Eqs.(\ref{eqn:ench}) and (\ref{eqn:sdensp}) that
the finite-size
spectrum of the {\it s-d} exchange model is
essentially identical to
that of the Anderson model under $\delta_F=\pi/2$.
Our results are
in agreement with those in \cite{af1}
for the {\it s-d} exchange
model.

\section{Summary}

We have studied the finite-size corrections
to the energy spectrum
for the Kondo problem based on the
Anderson model, the degenerate Anderson model and
the {\it s-d} exchange model. All these models are known to
exhibit the local
Fermi-liquid properties at low energies.  The basic picture
of local Fermi liquid was already established
and confirmed by the exact evaluation of
various static quantities.
A recent CFT approach to the Kondo problem by Affleck and Ludwig has
clarified how the Fermi-liquid picture emerges
and is described in terms of boundary CFT.
Our calculations demonstrate that the finite-size
spectrum consistent with  boundary CFT is derived directly
from the {\it microscopic} models. We have also pointed out
that  the critical behavior related
to the $X$-ray absorption singularities
will also be described  by boundary CFT.
A natural extension of the present work is to analyze
the multichannel Kondo problem including the
overscreened case. It is in principle possible to work out
the finite-size spectrum using the
Bethe-ansatz solution.  We hope to turn
to this issue in the near future.

\acknowledgments{}
The work of N. K. is supported in part by the Monbusho International
Scientific Research Program.
The work of S.-K.Y. is supported in part by Grant-in-Aid for
Scientific Research on Priority Area 231 ``Infinite Analysis''.

\newpage

\appendix{Finite-size corrections to
the ground-state energy of the
Anderson model}

We present the details of computations of
the finite-size corrections to the ground-state energy
of the Anderson model in Sec.I.
Applying the Euler-Maclaurin formula to Eqs.(\ref{eqn:bethean1})
and (\ref{eqn:bethean2}) one obtains the finite-size form of the
Bethe-ansatz integral equations. The result reads
\begin{eqnarray}
\rho_L (k)&=&{1\over 2\pi}+{1\over L}\Delta (k)+
B^{'}(k)\int^{\infty}_{\Lambda^-}{\rm d}\Lambda a_1(B(k)-\Lambda)
\sigma_L(\Lambda) \nonumber \\
& &+\frac{1}{24L^2}\biggl\{\frac{B^{'}(k)a_1^{'}(B(k)-\Lambda^{+})}
{\sigma_h(\Lambda^{+})}-\frac{B^{'}(k)a_1^{'}(B(k)-\Lambda^{-})}
{\sigma_h(\Lambda^{-})}\biggr\},
\label{eqn:bef1}
\end{eqnarray}
\begin{eqnarray}
\sigma_L(\Lambda)&+&\int_{\Lambda^-}^{\infty}{\rm d}\Lambda^{'}
a_2(\Lambda-\Lambda^{'})\sigma_L (\Lambda^{'})
+\frac{1}{24L^2}\biggl\{\frac{a_2^{'}(\Lambda-\Lambda^{+})}
{\sigma_h(\Lambda^{+})}-\frac{a_2^{'}(\Lambda-\Lambda^{-})}
{\sigma_h(\Lambda^{-})}\biggr\} \nonumber \\
&+&\int_{-\infty}^{k^{+}}{\rm d}k a_1(\Lambda -B(k))\rho_L (k)
+\frac{1}{24L^2}\biggl\{\frac{B^{'}(k^+)a_1^{'}(\Lambda-B(k^{+}))}
{\rho_h(k^{+})}-\frac{B^{'}(k^{-})a_1^{'}(\Lambda-B(k^{-}))}
{\rho_h(k^{-})}\biggr\} \nonumber \\
&=&A(\Lambda)+{1\over L}Z(\Lambda),
\label{eqn:bef2}
\end{eqnarray}
with $k^{+}=\beta$ and $\Lambda^{-}=\alpha$. We let
$k^{-}\rightarrow -\infty$ as well as
$\Lambda^{+}\rightarrow +\infty$ at the end of the calculation.
The solutions of Eqs.(\ref{eqn:bef1})
and (\ref{eqn:bef2}) are written as
\begin{equation}
\rho_L(k)=\rho_h(k)
+\frac{1}{L}\rho_{imp}(k)+\frac{1}{24L^2}\biggl\{
\frac{\rho_1^{+}(k)}{\rho_h(k^{+})}
+\frac{\rho_1^{-}(k)}{\rho_h(k^{-})}
+\frac{\rho_2^{-}(k)}{\sigma_h(\Lambda^{-})}
+\frac{\rho_2^{+}(k)}{\sigma_h(\Lambda^{+})}
\biggr\},
\label{eqn:rh}
\end{equation}
\begin{equation}
\sigma_L(\Lambda)=\sigma_h(\Lambda)
+\frac{1}{L}\sigma_{imp}(\Lambda)
+\frac{1}{24L^{2}}\biggl\{
\frac{\sigma_1^{+}(\Lambda)}{\rho_h(k^{+})}
+\frac{\sigma_1^{-}(\Lambda)}{\rho_h(k^{-})}
+\frac{\sigma_2^{-}(\Lambda)}{\sigma_h(\Lambda^{-})}
+\frac{\sigma_2^{+}(\Lambda)}{\sigma_h(\Lambda^{+})}
\biggr\}.
\label{eqn:sig}
\end{equation}
Here $\rho_i^{\pm}(k)$ and $\sigma_i^{\pm}(\Lambda)$
($i=1, 2$) satisfy
\begin{equation}
\rho_i^{\pm} (k)=\rho_i^{0\pm}
+B^{'}(k)\int^{\infty}_{\alpha}{\rm d}\Lambda a_1(B(k)-\Lambda)
\sigma_i^{\pm}(\Lambda),
\label{eqn:befi1}
\end{equation}
\begin{equation}
\sigma_i^{\pm}(\Lambda)+\int_{\alpha}^{\infty}{\rm d}\Lambda^{'}
a_2(\Lambda-\Lambda^{'})\sigma_i^{\pm}
(\Lambda^{'})+\int_{-\infty}^{\beta}{\rm d}k a_1(\Lambda -B(k))
\rho_i^{\pm}(k)
=\sigma_i^{0\pm},
\label{eqn:befi2}
\end{equation}
where $\rho_1^{0\pm}=0$, $\rho_2^{0\pm}=
\pm B^{'}(k)a_1^{'}(B(k)-\Lambda^{\pm})$,
$\sigma_1^{0\pm}=\pm B^{'}(k^{\pm})a_1^{'}(B(k^{\pm})-\Lambda)$
and $\sigma_2^{0\pm}=\pm a_2^{'}(\Lambda^{\pm}-\Lambda)$.

Let us next apply the Euler-Maclaurin formula
to Eq.(\ref{eqn:regulated}).
We then obtain the finite-size form of the ground-state energy
as follows:
\begin{eqnarray}
E_0=&L&\Big\{ \int^{\beta}_{-\infty}{\rm d} k \varphi (k)k
\rho_L(k) + 2\int^{\infty}_{\alpha}{\rm d}\Lambda x(\Lambda)
\varphi (x(\Lambda))\sigma_L(\Lambda) \Big\} \\ \nonumber
&-&\frac{1}{24L\rho_h(k^{+})}
+\frac{2x^{'}(\Lambda^{-})}{24L\sigma_h(\Lambda^{-})}.
\end{eqnarray}
To derive this we have used the assumed property
 $\varphi(k^{-}), \varphi^{'}(k^{-})
\rightarrow 0 $ as $k^{-}\rightarrow -\infty$ and
$\varphi(x(\Lambda^{+})),\varphi^{'}(x(\Lambda^{+}))
\rightarrow 0 $ as $\Lambda^{+}\rightarrow +\infty$.
Using Eqs.(\ref{eqn:rh}) and (\ref{eqn:sig}) we express
the ground-state energy in the form
\begin{eqnarray}
E_0=L\varepsilon_0
&+&\frac{1}{24L\sigma_h(\Lambda^{-})}
\biggl\{2x^{'}(\Lambda^{-})+\int^{\beta}_{-\infty}
{\rm d}k\varphi(k)k\rho_2^{-}(k)
+2\int^{\infty}_{\alpha}{\rm d}\Lambda
\varphi(x(\Lambda))x(\Lambda)\sigma_2^{-}(\Lambda)\biggr\}\nonumber
\\
&+&\frac{1}{24L\sigma_h(\Lambda^{+})}
\biggl\{\int^{\beta}_{-\infty}
{\rm d}k\varphi(k)k\rho_2^{+}(k)
+2\int^{\infty}_{\alpha}{\rm d}\Lambda
\varphi(x(\Lambda))x(\Lambda)\sigma_2^{+}(\Lambda)\biggr\}
\nonumber \\
&+&\frac{1}{24L\rho_h(k^{+})}
\biggl\{-1+\int^{\beta}_{-\infty}
{\rm d}k\varphi(k)k\rho_1^{+}(k)
+2\int^{\infty}_{\alpha}{\rm d}\Lambda
\varphi(x(\Lambda))x(\Lambda)\sigma_1^{+}(\Lambda)\biggr\}
\nonumber \\
&+&\frac{1}{24L\rho_h(k^{-})}
\biggl\{\int^{\beta}_{-\infty}
{\rm d}k\varphi(k)k\rho_1^{-}(k)
+2\int^{\infty}_{\alpha}{\rm d}\Lambda
\varphi(x(\Lambda))x(\Lambda)\sigma_1^{-}(\Lambda)\biggr\},
\label{eqn:apene}
\end{eqnarray}
where
\begin{equation}
\varepsilon_0=
\int^{\beta}_{-\infty}{\rm d}k \varphi (k)k
\rho(k)+ 2\int^{\infty}_{\alpha}{\rm d}\Lambda x(\Lambda)
\varphi (x(\Lambda))\sigma(\Lambda).
\end{equation}
It is easily seen from Eqs.(\ref{eqn:befi1}) and (\ref{eqn:befi2})
that the second and fourth lines of (\ref{eqn:apene})
vanish in the limit $k^{-}\rightarrow -\infty$,
$\Lambda^{+}\rightarrow +\infty$.
Furthermore the first and third lines can be expressed
in terms of the dressed energy functions (\ref{eqn:denes}) and
(\ref{eqn:denec})
\begin{equation}
E_0=L\varepsilon_0
+\frac{1}{24L\sigma_h(\alpha)}
\frac{\partial \varepsilon_c(\Lambda)}{\partial \Lambda}
\biggl\vert_{\Lambda=\alpha}\biggr.
+\frac{1}{24L\rho_h(\beta)}
\frac{\partial \varepsilon_s(k)}{\partial k}
\biggl\vert_{k=\beta}\biggr. .
\end{equation}
Here we note the relations
\begin{equation}
\frac{1}{\sigma_h(\alpha)}\frac{\partial
\varepsilon_c(\Lambda)}{\partial \Lambda}
\biggl\vert_{\Lambda=\alpha}\biggr.=-2\pi,\quad
\frac{1}{\rho_h(\beta)}\frac{\partial
\varepsilon_s(k)}{\partial k}
\biggl\vert_{k=\beta}\biggr.=-2\pi,
\end{equation}
which can be checked from
Eqs.(\ref{eqn:be1}), (\ref{eqn:be2}), (\ref{eqn:denes})
and (\ref{eqn:denec}).
Thus we arrive at Eq.(\ref{eqn:andcasimir}) in the text.

\newpage

\end{document}